\documentclass[prd,nofootinbib,showpacs]{revtex4}

\usepackage{graphicx}
\usepackage[usenames,dvipsnames]{color}
\usepackage{amsmath,amssymb}
\usepackage{epstopdf}
\usepackage{hyperref}

\begin{document}

\title{Spontaneous CP violation in quark scattering 
from QCD $Z\left(3\right)$ interfaces}

\author{Abhishek Atreya}
\email{atreya@iopb.res.in}
\affiliation{Institute of Physics, Bhubaneswar, 751005, India}
\author{Anjishnu Sarkar}
\email{anjishnu@prl.res.in}
\affiliation{Physical Research Laboratory, Ahmedabad, 380009, India}
\author{Ajit M. Srivastava}
\email{ajit@iopb.res.in}
\affiliation{Institute of Physics, Bhubaneswar, 751005, India}

\begin{abstract}
In this paper, we explore the possibility of spontaneous CP violation in the 
scattering of quarks and anti-quarks from QCD $Z\left(3\right)$ domain walls. 
The CP violation here arises from the nontrivial profile of the background 
gauge field $(A_{0})$ between different $Z\left(3\right)$ vacua. We 
calculate the spatial variation of $A_{0}$ across the Z(3) interface
from the profile of the Polyakov loop $L(\vec{x})$ for the
Z(3) interface and calculate the reflection of quarks and antiquarks using
the Dirac equation. This spontaneous CP violation has interesting 
consequences for the relativistic heavy-ion collision experiments, such 
as baryon enhancement at high $P_{T}$. It also acts as a source of 
additional J/$\psi$ suppression. We also discuss its implications 
for the early universe.
\end{abstract}

\maketitle

\section{Introduction}
\label{sec:intro} 

 The possibility of extended topological objects in the quark-gluon plasma
(QGP) phase, e.g. Z(3) interfaces arising from spontaneous breaking of
Z(3) symmetry, has been extensively discussed in the literature 
\cite{Bhattacharya:1992qb,West:1996ej,Boorstein:1994rc}. 
It has also been pointed out that there are also topological string
defects in QGP forming at the junctions of Z(3) walls \cite{Layek:2005fn}.
Formation and evolution of these objects in the initial transition 
to the QGP phase has been studied in the context of relativistic heavy-ion 
collision experiments (RHICE) \cite{Gupta:2010pp}. Certain consequences of
Z(3) walls for baryon inhomogeneity generation in the universe have 
also been explored \cite{Layek:2005zu}. 
Investigation of these objects is important not only for probing
the very rich vacuum structure of the QCD in the deconfining phase, but
also because these provide the only example of topological defects
in a relativistic quantum field theory which can be probed in laboratory
conditions, namely, the relativistic heavy-ion collision experiments
(RHICE). The existence of these objects has been questioned in the 
literature, especially in the presence of quarks \cite{Smilga:1993vb,
Belyaev:1991cw}. However, there are recent Lattice studies by Deka et al. 
\cite{Deka:2010bc} of QCD with quarks which have attempted to directly probe 
the existence of different Z(3) vacua. These results show strong possibility 
of the existence of non-trivial, metastable, Z(3) vacua for high temperatures.
The exact value of the temperature, above which these metastable Z(3)
vacua are seen, is not important. What is important is that these vacua 
seem to exist as metastable thermodynamic phases of QCD in the deconfining
regime, and hence associated topological objects will necessarily
arise in any realistic phase transition from the confining phase
to the QGP phase. 

  In this paper we will investigate an interesting possibility
arising from the existence of Z(3) interfaces. We will study reflection
of quarks and antiquarks from Z(3) walls and show the existence of
CP violation arising from the Z(3) walls. This CP violation is
spontaneous, arising due to the background configuration of the
gauge field corresponding to the Z(3) wall, and was first demonstrated
by Altes et al. \cite{KorthalsAltes:1994if}. It was shown in ref. 
\cite{KorthalsAltes:1994if}, in the context of the universe, that due to the 
non-trivial background field configuration for the standard model gauge 
fields, the localization of quarks and antiquarks on the wall is different. 
Its possible effects on the electroweak baryogenesis via sphalerons was 
discussed in \cite{KorthalsAltes:1994if}. Same possibility of spontaneous CP 
violation for the case of QCD was also discussed in 
\cite{KorthalsAltes:1992us}.    
We extend these studies by calculating the propagation of quarks and
antiquarks across the Z(3) walls and show that they have different 
reflection coefficients. For this we calculate the profile of the
order parameter $L\left(\vec{x}\right)$ between different Z(3) vacua 
\cite{Layek:2005fn}  using the the effective potential for the Polyakov 
loop, as proposed by Pisarski \cite{Pisarski:2000eq}. We then obtain
obtain the profile of the background gauge field $A_0$ from this
$L\left(\vec{x}\right)$ profile. This $A_0$ configuration provides a 
potential for the propagation of quark causing non-trivial reflection of
quarks from the wall. It is important to know the uncertainties in the 
determination of the $A_0$ profile depending on the choice of the
specific form of the effective potential, such as those given in
\cite{Fukushima:2003fw,Roessner:2006xn}. To address this issue we
repeat the above calculation for another choice of effective potential
of the Polyakov loop as provided by Fukushima \cite{Fukushima:2003fw}. We find
that, even though the two effective potentials (in refs. 
\cite{Pisarski:2000eq} and  \cite{Fukushima:2003fw}) are of qualitatively 
different shapes, the resulting wall profile and $A_0$ profile are surprisingly
similar. This gives us confidence in the use of our procedure to
calculate the reflection of quark and antiquarks from the Z(3) interfaces. 

 Different values of the reflection coefficients of quarks and antiquarks
from the Z(3) walls will have very interesting implications
for the case of RHICE and for the early universe. Here we mention that
in the earlier studies by some of us the reflection of quarks/antiquarks 
from Z(3) walls (in the context of RHICE and the universe) \cite{Layek:2005zu}, 
was studied by modeling the dependence of effective quark mass on the 
magnitude of the Polyakov loop, and no possibility of spontaneous CP 
violation was explored.
This CP violation, resulting in different reflection coefficients of
quarks and antiquarks from Z(3) walls, will lead to segregation of quarks 
and anti-quarks due to motion (collapse) of walls. As a result  there will
be selective concentration of baryon (or antibaryon) number in different 
regions, depending on the Z(3) vacua involved. This will have direct 
observable consequences for the relativistic heavy ion collision experiments. 
For example, it will affect the yield of baryons and mesons, 
enhancing baryon multiplicities and suppressing meson multiplicities. 
As we will see, these effects are expected to be important for heavy
quarks, especially for charm and heavier flavors. A detailed analysis of 
these effects is planned for a future work. This CP violation can also play an 
important role in the context of early universe, especially for generation
of baryon density inhomogeneities, by 
segregating baryons and antibaryons. We mention here that our analysis of 
reflection of quarks in this paper utilizes Z(3) wall profile of pure SU(3) 
gauge theory, without dynamical quarks. The effects of quarks may 
not be important in the context of RHICE due to small length and time scales 
involved, but for the case of universe these effects will be of crucial 
importance. We will discuss this further below. 

  The paper is organized in the following manner. In section II
we discus the basic physics of the origin of spontaneous CP violation
due to the presence of Z(3) interfaces \cite{KorthalsAltes:1992us,
KorthalsAltes:1994if} and discuss the effective potential for the Polyakov 
loop, as proposed by Pisarski \cite{Pisarski:2000eq} for calculating 
various quantities. In section III, we discuss how to obtain the profile 
of the background gauge field $A_0$ from the profile of the order 
parameter $L\left(\vec{x}\right)$ between different Z(3) vacua 
\cite{Layek:2005fn}. In section IV we address the issue of uncertainties 
in the determination of the $A_0$ profile depending on the choice of the
specific form of the effective potential by repeating the calculations of
section III for the effective potential of the Polyakov loop provided by 
Fukushima \cite{Fukushima:2003fw}. The resulting wall profile and $A_0$
profile are found to be very close to those found in section III.  We use the 
profile of $A_0$ as calculated in section III, for the Dirac equation 
(in the Minkowski space) in section V to calculate the reflection and 
transmission coefficients for quarks and antiquarks. Section VI presents 
our results and conclusions are discussed in section VII. 

\section{ORIGIN OF SPONTANEOUS CP VIOLATION}

 We first discuss the basic physics of the origin of the spontaneous
CP violation from the existence of Z(3) walls. For the case of pure
$SU(N)$ gauge theory, we start with the definition of the Polyakov loop,
\cite{Polyakov:1978vu,Gross:1980br,McLerran:1981pb}

\begin{equation} 
L(x) = \frac{1}{N}Tr\biggl[\mathbf{P} \exp\biggl(ig\int_{0}^{\beta}A_{0}
  (\vec{x},\tau)d\tau\biggr)\biggr],
\label{eq:lx}
\end{equation}

where, $A_{0}(\vec{x},\tau) = A_{0}^{a}(\vec{x},\tau)T^{a}, (a = 1,\dotsc N)$
are the gauge fields and $T^{a}$ are the generators of $SU\left(N\right)$
in the fundamental representation. $\mathbf{P}$ denotes the path ordering in
the Euclidean time $\tau$, and $g$ is the gauge coupling. Under global 
$Z(N)$ symmetry transformation, the Polyakov Loop transforms as
\begin{equation}
L(x) \longrightarrow Z\times L(x), \qquad \textrm{where } Z = e^{i\phi}.
\end{equation}
Here, $\phi = 2\pi m/N$; $m = 0,1 \dotsc (N-1)$. 

Thermal average of the Polyakov loop, $\langle L(x)\rangle$, is the order 
parameter for the 
confinement-deconfinement phase transition. (From now onwards, we will
use $L(x)$ to denote $\langle L(x)\rangle$,) It is related to the free 
energy of a test quark in a pure gluonic medium ($L(x) \propto e^{-\beta F}$). 
$L(x) \neq 0$ implies finite free energy of a test quark and hence, 
the deconfined phase (i.e the system  is above the critical temperature 
$T_{c}$). This leads to spontaneous breaking of $Z(N)$ symmetry. On the 
other hand, $L(\vec{x}) = 0$ implies infinite free 
energy of a test quark and hence, confined phase (i.e. system is below 
$T_{c}$). The $Z(N)$ symmetry is then restored. The $N$-fold degeneracy 
of the ground state implies the existence of interfaces between regions of
different $Z(3)$ vacua. For QCD, the gauge group is the color group 
$SU(3)_{c}$. It has three $Z(3)$ vacua resulting from the spontaneous breaking 
of $Z(3)$ symmetry in the high temperature (deconfined) phase characterized 
by,
\begin{equation}
L(\vec{x}) = 1, e^{i2\pi/3}, e^{i4\pi/3}.
\end{equation}
 As we mentioned above, there have been questions whether these Z(3)
domains have some physical meaning or not 
\cite{Smilga:1993vb,Belyaev:1991cw}. The inclusion of quarks raises 
further issues as they do not respect the $Z(N)$ symmetry. It has been
argued that it is possible to interpret the effect of addition of quarks as 
the explicit breaking of $Z(N)$ symmetry and lifting of degeneracy of the 
vacuum \cite{Pisarski:2000eq,Dumitru:2000in,Dumitru:2002cf,Dumitru:2001bf},
and we will follow this approach. Further, as we mentioned in the 
Introduction, recent lattice QCD studies with quarks \cite{Deka:2010bc}
have strengthened the physical basis for
the existence of these different Z(3) vacua. The metastability of
non-trivial $Z(3)$ vacua will have important implications for RHICE and the 
early universe. However, for the rest of the paper we will consider the 
pure gauge case for calculating the $Z(3)$ interface profiles. This is because 
our main objective here is to show the interesting possibility of spontaneous 
CP violation in the reflection of quarks and antiquarks from Z(3) walls which 
is independent of the explicit symmetry breaking. We will briefly comment
on the effects of quarks in the last section, detailed study of these
effects will be presented in a future work.
 
As mentioned earlier, different $Z(3)$ vacua have interpolating
$L(\vec{x})$ profile leading to Z(3) interfaces. This essentially means 
that there is a background gauge field $A_{0}(\vec{x})$ profile which 
interpolates between different $Z(3)$ vacua. The quarks/anti-quarks moving 
across the $Z(3)$ domain walls will behave differently in the presence of a 
given spatially varying $A_{0}$ field configuration. As a result,
we should have different reflection and transmission coefficient for 
quarks and anti-quarks. This is the source of CP violation. The origin of
this CP asymmetry is spontaneous in nature.  The earlier studies 
\cite{KorthalsAltes:1992us,KorthalsAltes:1994if}
of this spontaneous CP violation arising from Z(3) walls focused on the 
localized solution of Dirac equation (in Euclidean space), and it was shown 
that if a wave function for a fermion  species localizes, then it's CP 
conjugate doesn't. The whole discussion in ref. \cite{KorthalsAltes:1992us,
KorthalsAltes:1994if} was within the Euclidean formalism and the exact gauge 
field profile was not determined in these investigations.

In this paper we are interested in the calculation of reflection and 
transmission coefficient of quarks and anti-quarks and hence, in the 
propagating solutions. It is important to note here that the background 
gauge field profile comes from the finite temperature field 
theory, which is formulated in the Euclidean space. To calculate the reflection
and the transmission coefficients (or to study propagation of quarks, in 
general), we need to solve Dirac equation in the Minkowski space. 

We start with the Dirac equation in the Euclidean space, with the spatial
dependence of $A_0$ calculated from Z(3) wall profile as mentioned above. 
Then we do the analytic continuation of the full equation to the Minkowski 
space and use the resulting equation to calculate the reflection and 
transmission coefficients. We should mention here that it may seem
puzzling that we are extracting information about colored objects
(i.e. $A_0$) starting with a colorless object, the Polyakov loop. However,
as we will explain later in Section V, starting with a given profile
of $L(x)$, one does not get unique solution for $A_0(x)$ and the ambiguity
about color information manifests itself in the form of a set of 
solutions of $A_0$.

 We will use the effective model for the Polyakov
loop as proposed by Pisarski \cite{Pisarski:2000eq}. The Lagrangian density 
has the form
\begin{equation} 
\mathcal{L} = \frac{N}{g^{2}}\vert \partial_{\mu}L\vert^{2}T^{2} - V(L).
\label{eq:lagrangian}
\end{equation}
$N=3$ for our case (i.e QCD). $T^{2}$ is multiplied with the first term to
give the correct dimensions to the kinetic term. 
$V(L)$ is the potential term that has the form

\begin{equation}
V(L) = \biggl(-\frac{b_2}{2}|L|^{2} - \frac{b_3}{6}\Bigl(L^{3} 
+ (L^{*})^{3}\Bigr) + \frac{1}{4}(|L|^{2})^{2}\biggr)b_4T^{4}.
\label{eq:pispot}
\end{equation}

The cubic term in $L(\vec{x})$ in the above potential, when written in 
terms of $L(\vec{x}) = |L(x)| e^{i\theta}$, gives rise
to $\cos(3\theta)$ term that leads to three degenerate $Z(3)$ vacua when 
$L(\vec{x}) \neq 0$ (i.e. when $T > T_{c}$). The coefficients $b_{2}$, 
$b_{3}$ and $b_{4}$, in the potential, are fixed in ref \cite{Dumitru:2000in,
Dumitru:2002cf,Dumitru:2001bf} by comparing 
with lattice results for the pressure and energy density for pure SU(3) 
gauge theory \cite{Boyd:1996bx,Okamoto:1999hi}.
$b_{2}$ is given by $b_2 = \left(1-1.11/x\right) 
\left(1+0.265/x\right)^{2}\left(1+0.300/x\right)^{3} - 0.478$, where 
$x=T/T_{c}$ with $T_{c}\sim182$ MeV. The other parameters are $b_3 = 2.0$ 
and $b_4 = 0.6061\times47.5/16$ (the factor 47.5/16 for $b_4$ is to account
for the additional quark degrees of freedom compared to pure SU(3) case).  
With the above values, $L\left(\vec{x}
\right) \longrightarrow y = b_{3}/2 +\frac{1}{2} \times \sqrt{b_{3}^{2} 
+ 4b_{2}\left(T=\infty\right)}$ as $T \longrightarrow \infty$. $L\left(\vec{x}\right)$ and other quantities are normalized as follows,
\begin{equation}
L\left(\vec{x}\right) \longrightarrow L\left(\vec{x}\right)/y, 
\qquad b_{2} \longrightarrow b_{2}/y^{2}, \qquad b_{3}
\longrightarrow b_{3}/y, \qquad b_{4} \longrightarrow b_{4}y^{4},
\end{equation}
so that $L\left(\vec{x}\right) \longrightarrow
1$ as $T \longrightarrow \infty$. The normalized quantities are then used 
in eqn. (\ref{eq:pispot}), which is then used to calculate the 
$L\left(\vec{x}\right)$ profile using energy minimization, see 
ref.\cite{Layek:2005fn} for details. Fig.1 shows the plot of $|L({\vec x})|$
for the interface between two different vacua (in the absence of quarks
all the three interfaces have same profile for $|L({\vec x})|$). 
We mention that the surface tension $\sigma$ of the Z(3) walls was estimated
in refs.\cite{Layek:2005zu} for the above effective potential and
it was found that $\sigma = $ 0.34, 2.62, and 7 GeV/fm$^2$ for
$T = $ 200, 300, and 400 MeV respectively. There have been Lattice studies
of Z(3) wall tension. In ref.\cite{Kajantie:1990bu} the surface tension was
found to be $\sigma(T_c) =  0.17 T_c^3$. With $T_c = 182$ MeV
the $T = 200$ result for $\sigma$ in ref.\cite{Layek:2005zu} is 
larger by almost factor 10 than the lattice result of
ref.\cite{Kajantie:1990bu}.
However, the values of $\sigma$ for larger temperatures, $T$ = 300 and 
400 MeV are in reasonable agreement with the analytical estimates 
\cite{Bhattacharya:1990hk} (which give $\sigma = {4(N-1)\pi^2 T^3
\over 3\sqrt{3} g}$ for large temperatures). 

\begin{figure}
\begin{center}
\includegraphics[width=0.25\textwidth]{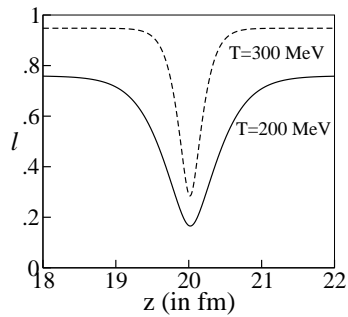}
\caption{Variation of $|L\left(\vec{x}\right)|$  between different $Z(3)$ vacua 
for $T=200$ MeV and $T = 300$ MeV respectively, as a function of $z$. Note 
that at higher temperature, the wall thickness is smaller, as expected.}
\label{fig:lprfl}
\end{center} 
\end{figure}

The energy minimization program gives the full profile for $L({\vec x})$
which is then used for calculating $A_0({\vec x})$ as described in the
next section. (As we mentioned in the Introduction, we will also consider
another form of effective potential as provided by Fukushima 
\cite{Fukushima:2003fw} in section IV.) 

\section{Obtaining $A_{0}$ profile}
\label{sec:a0cal}

In this section we calculate the $A_{0}$ profile form $L(\vec{x})$ profile by 
inverting eqn.(\ref{eq:lx}). As in ref. 
\cite{KorthalsAltes:1994if} we choose $A_{0}$ to be of the form
\begin{equation} 
A_{0} = \frac{2\pi T}{g}\left(a\lambda_{3} + b\lambda_{8}\right),
\label{eq:a0diag}
\end{equation}
where, $g$ is the coupling constant and $T$ is the temperature, while 
$\lambda_{3}$ and $\lambda_{8}$ are the diagonal Gell-Mann matrices. 
Coefficients $a$ and $b$ depend only on spatial coordinates. The advantage 
of taking this gauge choice is that we are dealing with the 
eigenvalues of the matrices that are invariant under gauge transformation. 

 We take $A_0$ to be independent of $\tau$. This is for simplicity. Further,
it can be justified in the high temperature limit due to periodic boundary
conditions on $A_0$ in the (Euclidean) time direction in the imaginary
time formalism being used here for finite temperature field theory. 

Substituting eqn.(\ref{eq:a0diag}) in eqn. (\ref{eq:lx}), we get

\begin{equation}
3L(x) = \exp (i\alpha) + \exp(i\beta) + \exp(i\gamma),
\label{eq:lsimplify}
\end{equation}

where, $\alpha = 2\pi\left(\frac{a}{3} + \frac{b}{2}\right)$ ,
$\beta = 2\pi\left(\frac{a}{3} - \frac{b}{2}\right)$ and $\gamma
= 2\pi(\frac{-2a}{3})$. On comparing the real and imaginary part of
eqn. (\ref{eq:lsimplify}), we get

\begin{subequations}
\begin{align}
\cos\left(\alpha\right) + \cos \left(\beta\right)
    + \cos \left(\gamma\right)& = 3 |L| \cos \left(\theta \right), \\
\sin\left(\alpha\right) + \sin \left(\beta\right)
    + \sin \left(\gamma\right)& = 3 |L| \sin \left(\theta \right).
\end{align}
\label{eq:lgrp}
\end{subequations}

 Here $\theta$ is defined by writing $L(x) = |L(x)| e^{i\theta}$.
In eqn. (\ref{eq:lx}), $A_{0}$ appears in the phase, so any increment in
the phase by a factor of type $2\pi n$ will result in the
same value of $L\left(\vec{x}\right)$. We first consider the above 
equations for $L = 1$ vacuum. Note that $|L| < 1$ for finite temperatures. 
However, we will keep referring to the three Z(3) vacua as $L =1, Z, Z^2$. The 
solutions are a set of ordered pairs $\left(a,b\right)_{L=1}$. These different 
solution sets reflect $2\pi n$ ambiguity in $A_{0}$. Similarly, we find the 
solution sets $\left(a,b\right)_{L=Z}$ corresponding to the 
$L = Z = exp(i2\pi/3)$ vacuum. One now needs to find the appropriate values 
of $\left(a,b\right)$ for the entire profile of $L(x)$ interpolating
between these two vacua. One ambiguity in this is obvious. It may
appear that any of the sets $\left(a,b\right)_{L=1}$ could be matched to 
any of the sets $\left(a,b\right)_{L=Z}$ as all sets for a given vacua 
are equivalent. However, this could lead to different $A_0$ profiles in 
between, which in turn would lead to different reflection and transmission 
coefficients. This problem is resolved when we realize that the variation 
of $A_{0}$ should be smooth across the domain wall. Thus, we 
can simply start with any one pair $\left(a,b\right)_{L=1}$, and set it 
as the initial condition for the generation of the profile of $A_0$ as 
one traverses the wall starting from $L = 1$ vacuum to $L = exp(i2\pi/3)$ 
vacuum. We only require that $a$ and $b$ vary smoothly as the profile
of $L(x)$ changes smoothly across the wall. It will then automatically lead
to the appropriate values of $\left(a,b\right)_{L=Z}$ as $L = Z$ vacuum
is approached.  

\begin{figure}[!htp]
\begin{center}
\begin{tabular}{ccc}
\includegraphics[width=0.45\textwidth]{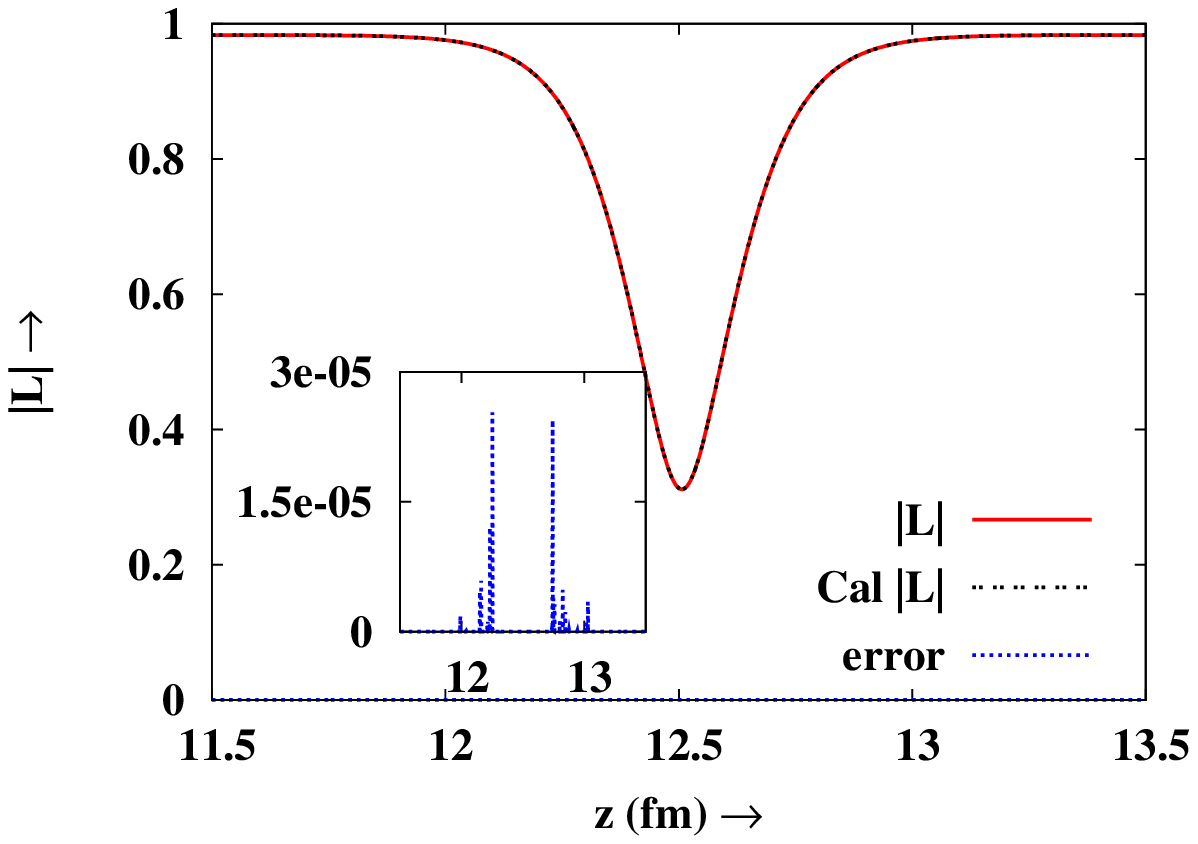}
&\qquad&
\includegraphics[width=0.45\textwidth]{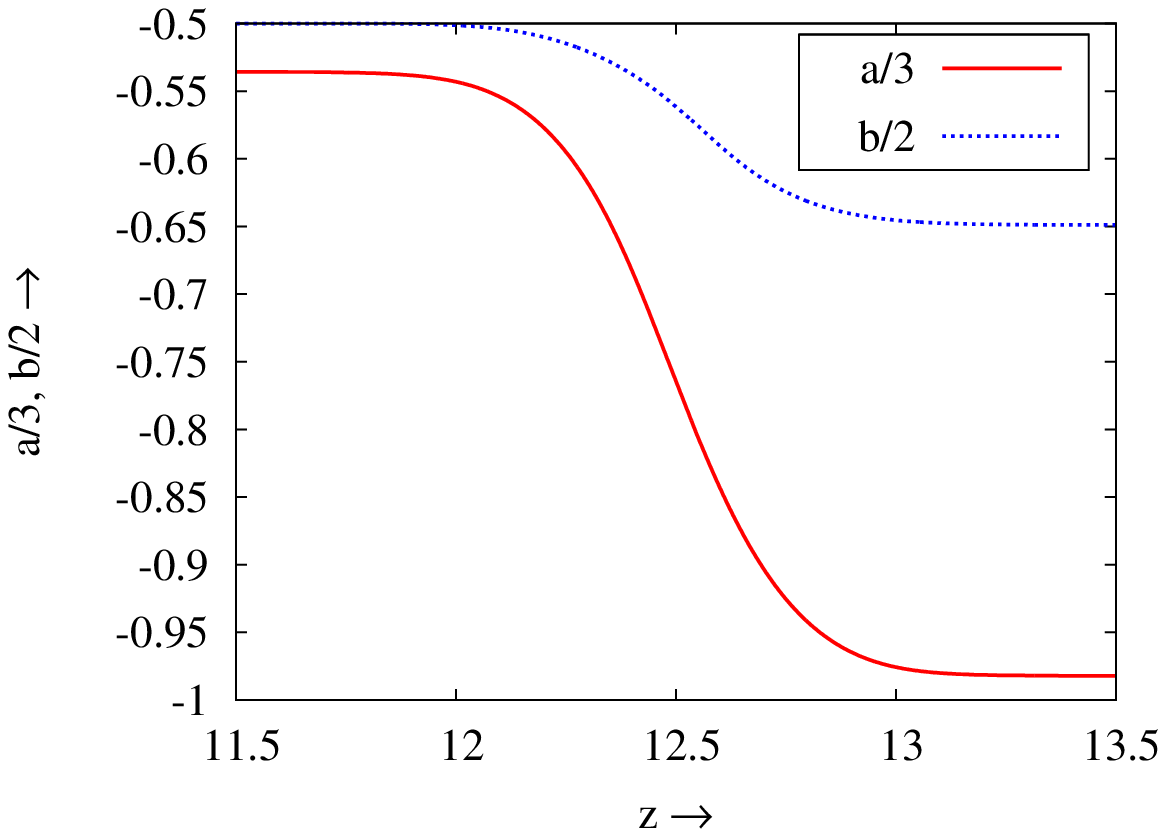} \\
(a) && (b)
\end{tabular} 
\caption{On left: Plot of calculated $|L|$ and the one obtained from 
minimizing the energy. The inset figure shows the deviation between the two 
profiles. On right: Variation of $a$ and $b$ between the regions $L(\vec{x}) =1$ and $L(\vec{x}) =e^{i2\pi/3}$. Initial point is $(-1.5,-1.0)$} 
\label{fig:lcalplt}
\end{center}
\end{figure}

 For the results shown here we had taken the initial values of
$\left(a,b\right) = \left(-1.5,-1.0\right)$ for the $L=1$ vacuum (in
a region far left to the interface). As one approaches the interface,
say, along the z axis, new value of $L(x)$ is selected from the profile
of $L(x)$ (calculated from the energy minimization program). We then take
small range of values near the original 
$\left(a,b\right) = \left(-1.5,-1.0\right)$ and 
$L(z)$ was then calculated for all these values. Those
values of $a$ and $b$ were selected for which the error between the
calculated $L$ and $L$ obtained by energy minimization was minimum. 
The process was then repeated for each value 
of $z$ to obtain $a,b$ values. Comparison between the calculated $|L|$ profile 
and the one obtained by energy minimization is given in figure 
(\ref{fig:lcalplt}a). It clearly shows that this technique works well.
Figure (\ref{fig:lcalplt}b) shows profile of parameters 
$a$ and $b$ across the domain wall.

\begin{figure}[!htp]
\begin{center}
\includegraphics[width=0.6\textwidth]{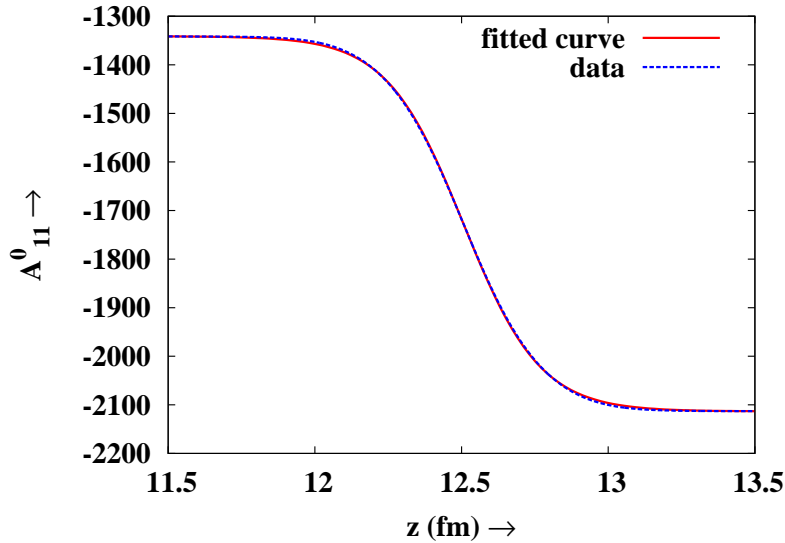}
\caption{Plot of calculated $A_{0}$ and the fitted profile ($A_{0}(x) = 
p \tanh (qx + r) + s$). The parameters have values $p = -378.27$, 
$q = 7.95001$, $r = -49.7141$, $s = -1692.48$. Only $(1,1)$ component of 
$A_{0}$ is plotted. The other components also have similar fit.}
\label{fig:a0plt}
\end{center}
\end{figure}

The calculated $a,b$ were then used to calculate $A_{0}$ using eqn 
(\ref{eq:a0diag}). The $A_{0}$ profile thus obtained is reasonably well 
fitted to the function $A_{0}(x) = p \tanh (qx + r) + s$ using 
gnuplot. The calculated $A_{0}$ profile and fitted
$A_{0}$ profile are plotted in figure (\ref{fig:a0plt}).

\section{Calculation of $A_0$ profile for a different effective potential}
 
 We now address the issue of the uncertainties in the determination of the 
$A_0$ profile depending on the choice of the specific form of the 
effective potential. Other parametrization of the effective 
potential for the Polyakov loop have been given in the literature,
e.g. in refs.\cite{Fukushima:2003fw,Roessner:2006xn}, and we will repeat
the calculations of the previous section for the effective 
potential of the Polyakov loop as provided by Fukushima 
\cite{Fukushima:2003fw}. 
For spatially varying $L$ configurations, we will continue to use the 
derivative terms as in Eq.(4) with general dimensional considerations 
(with suitable normalization of $L$). The effective potential for 
ref.\cite{Fukushima:2003fw} has the following form

\begin{equation} 
V[L]/T^4 = -2(d-1)e^{-\sigma a/T} |Tr L|^2 -ln[-|Tr L|^4 + 
8 Re(Tr L)^3 - 18|Tr L|^2 + 27]
\label{eq:lagrangian2}
\end{equation}
  $\sigma = $ (425 MeV)$^2$ is the string tension and
$2(d-1)e^{-\sigma a/T_d}$ = 0.5153 with $T_d = 270$ MeV is taken as
the transition temperature by choosing the lattice spacing $a = $
(272 MeV)$^{-1}$. Note that for consistency with the notations of Ref.[13], 
we will use $T_d$ and $T_c$ interchangeably, both meaning the deconfinement 
transition temperature. $L$ is the Polyakov loop but without the normalizing
factor of $N_c$ (= 3). (Thus, using with Eq.(4) we re-write the
above effective potential in terms of the normalized Polyakov loop. 
Henceforth by $L$ even for the above equation we will mean this
normalized Polyakov loop). It has been argued by Schaefer et al. 
\cite{Schaefer:2007pw} that the transition temperature has to be tuned 
depending on the number of quark flavors $N_f$ (and also the value of the 
baryon-chemical potential). In ref. \cite{Schaefer:2007pw}, the value of 
$T_d = 270$ MeV corresponds to the pure SU(3) case with $N_f = 0$.
In section II we have used the effective potential where the
coefficient $b_4$ is suitably normalized for the case of 3 flavors,
$N_f = 3$. For the case of $N_f = 3$, the value of transition temperature
from ref. \cite{Schaefer:2007pw} is $T_d$ = 178 MeV. Thus, we will use this
value of $T_d$ for the effective potential in Eq.(10).

 The effective potential in Eq.(10) is of qualitatively different nature
than the one given in Eq.(5). For small values of $L$ the two forms
will be similar as one can see by the expansion of the Logarithmic term
in the above equation. However, for  $|L|$ approaching
1 the two potentials are dramatically different. $V[l]$ in Eq.(10)
diverges at this limiting value thereby constraining 
$|L|$ within value 1. There is no such constraint in Eq.(5). Even the
shape of $V[L]$ is very different away from the origin, especially
near the three $Z(3)$ vacua. It is thus reasonable to expect that the
resulting profile of $Z(3)$ wall and resulting $A_0$ profile (using
calculations of previous sections) for Eq.(10) may be quite different 
from the ones obtained in section III for Eq.(5).

  With diverging $V[L]$ at  $|L| = 1$ in Eq.(10), and 
due to its non-trivial shape near the $Z(3)$ vacua, the application of
the technique of ref.\cite{Layek:2005fn} for the determination of $L$ profile
between two Z(3) vacua is much more complicated here. Especially 
non-trivial is the choice of initial ansatz for the wall profile 
which is used for the energy minimization program. In ref.\cite{Layek:2005fn},
the initial profile was taken to linearly interpolate between the two
Z(3) vacua as a function of spatial coordinate $z$. This choice
simply does not work for Eq.(10) due to the fact that $V[L]$ diverges
at  $|L|$ = 1 and linear interpolation takes it outside this bound.
For this we chose the initial trial profile to consist of two
parts, one  linearly decreasing  (with $z$) to $L = 0$ along $\theta = 0$ 
from the vacuum value and join this with the second part linearly 
increasing (with $z$) along $\theta = 2\pi/3$ to the second vacuum value.
This keeps the initial profile within the allowed region of $V[L]$
in Eq.(10).

 Second complication arises with the algorithm of energy minimization
itself. In ref.\cite{Layek:2005fn} correct $L$ profile was obtained
from the initial trial profile by fluctuating the value of $L$ at each
lattice point and determining the acceptable fluctuation which 
lowers the energy (with suitable overshoot criterion etc. as described
in detail in ref.\cite{Layek:2005fn}). However, with Eq.(10), fluctuations
of $L$ can take it out of the allowed region of $V[L]$. For this, we
skip those fluctuations which take $L$ outside the allowed region.
With these modification in the procedure, we were able to determine
the profile of the Z(3) wall and associated $A_0$ profile. In section III
we had calculated the profiles for temperature $T$ = 400 MeV (with
$T_c = $ 182 MeV for the effective potential in Eq.(5)). For the sake of
comparison with that case, for $V[L]$ in Eq.(10) with $T_c$ = 178 MeV
\cite{Schaefer:2007pw}, we calculate the profiles for $T = 391$ MeV which is
close enough to the value $T = 400 $ MeV, and has the same value for 
$T/T_c$.

\begin{figure}[!htp]
\begin{center}
\begin{tabular}{ccc}
\includegraphics[width=0.45\textwidth]{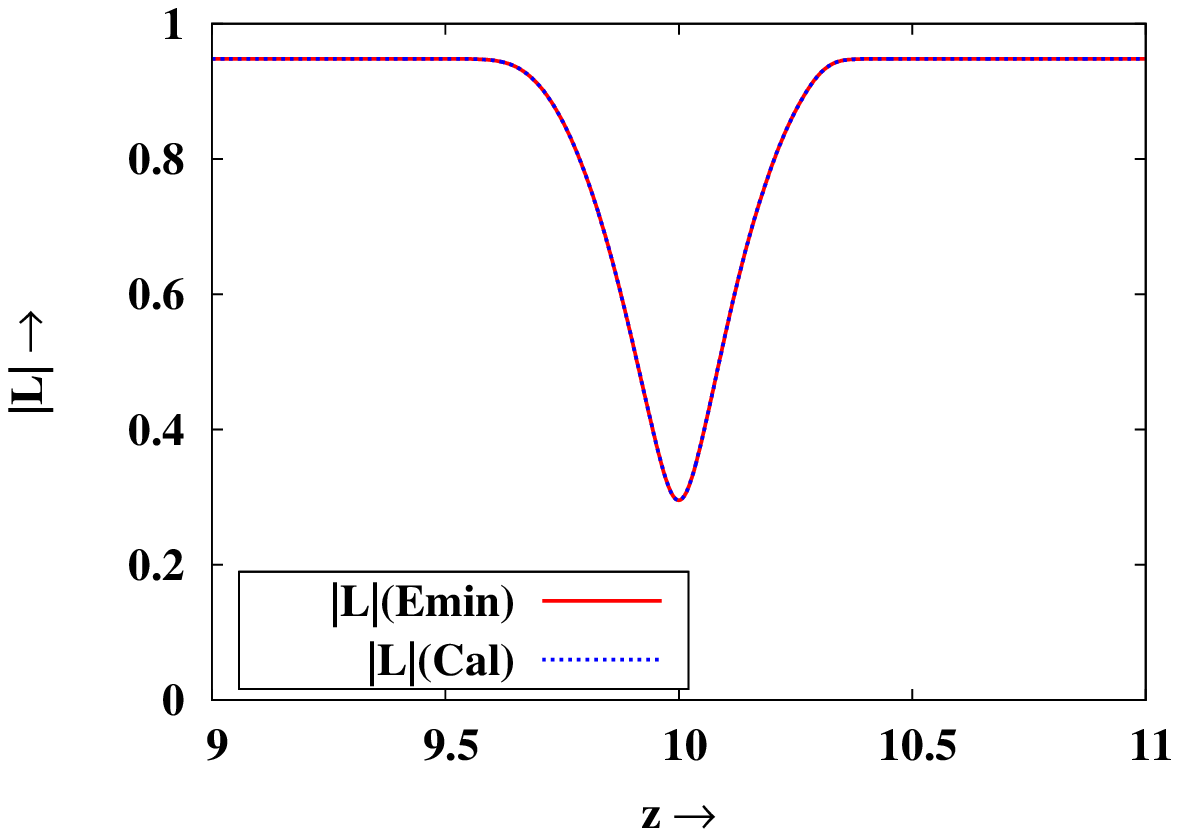}
&\qquad&
\includegraphics[width=0.45\textwidth]{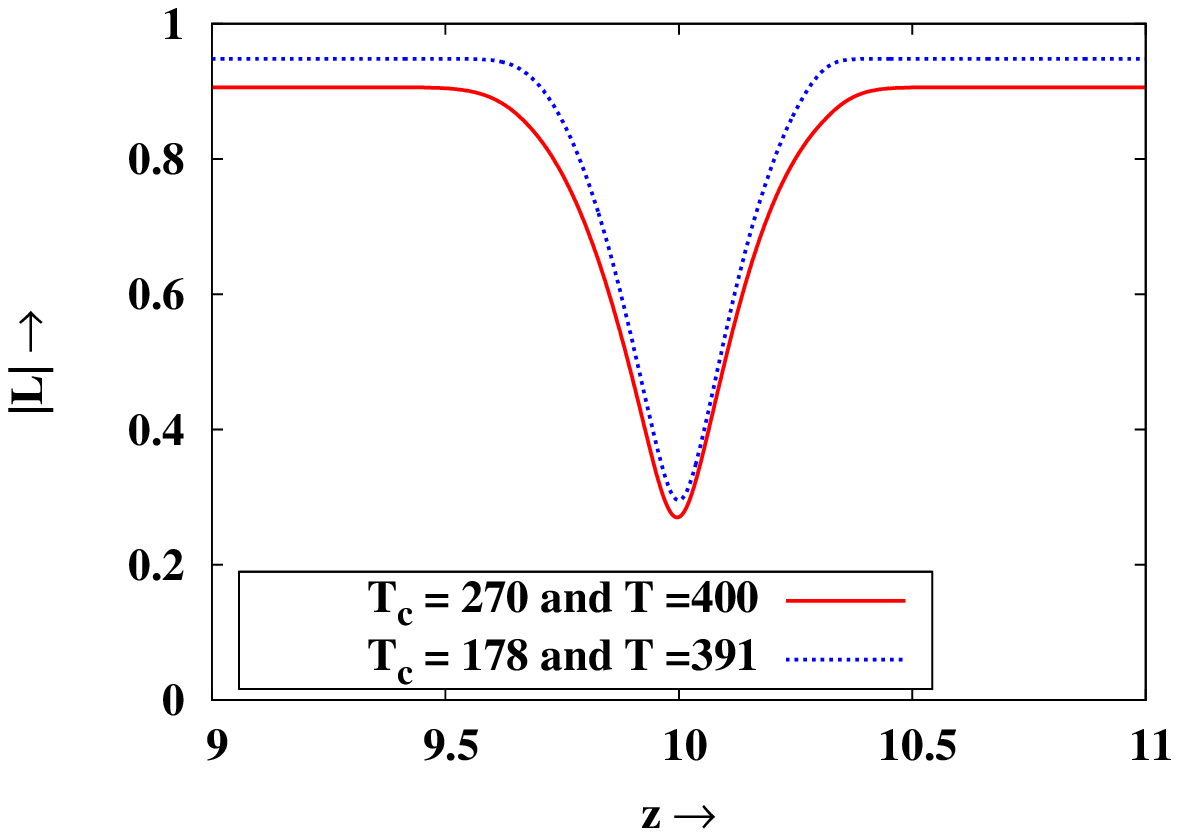} \\
(a) && (b)
\end{tabular} 
\caption{(a) Plot of the profile of $|L|$ corresponding to the 
effective potential in Eq.(10). (b) Comparison of the profiles of 
$|L|$ for different choices of $T_d$ in Eq.(10).}
\label{fig:lcalfkshm}
\end{center}
\end{figure}

\begin{figure}[!htp]
\begin{center}
\begin{tabular}{ccc}
\includegraphics[width=0.45\textwidth]{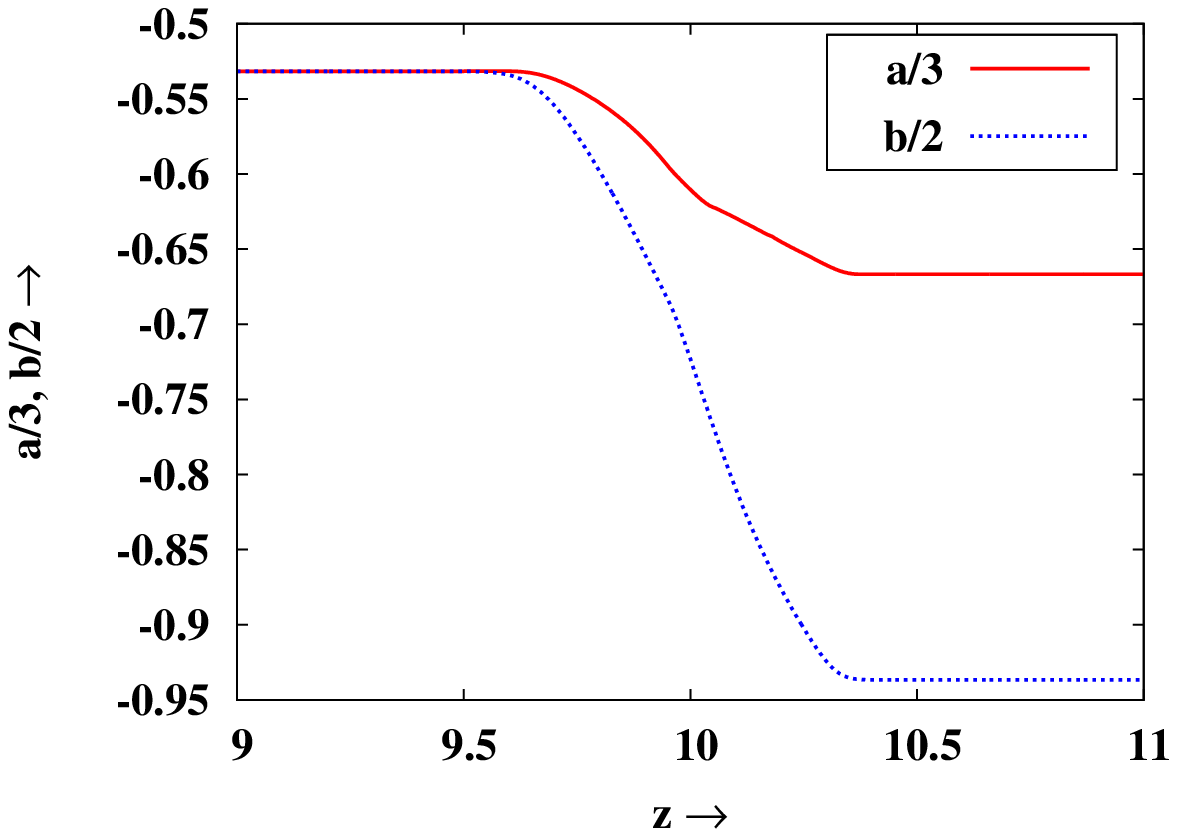}
&\qquad&
\includegraphics[width=0.45\textwidth]{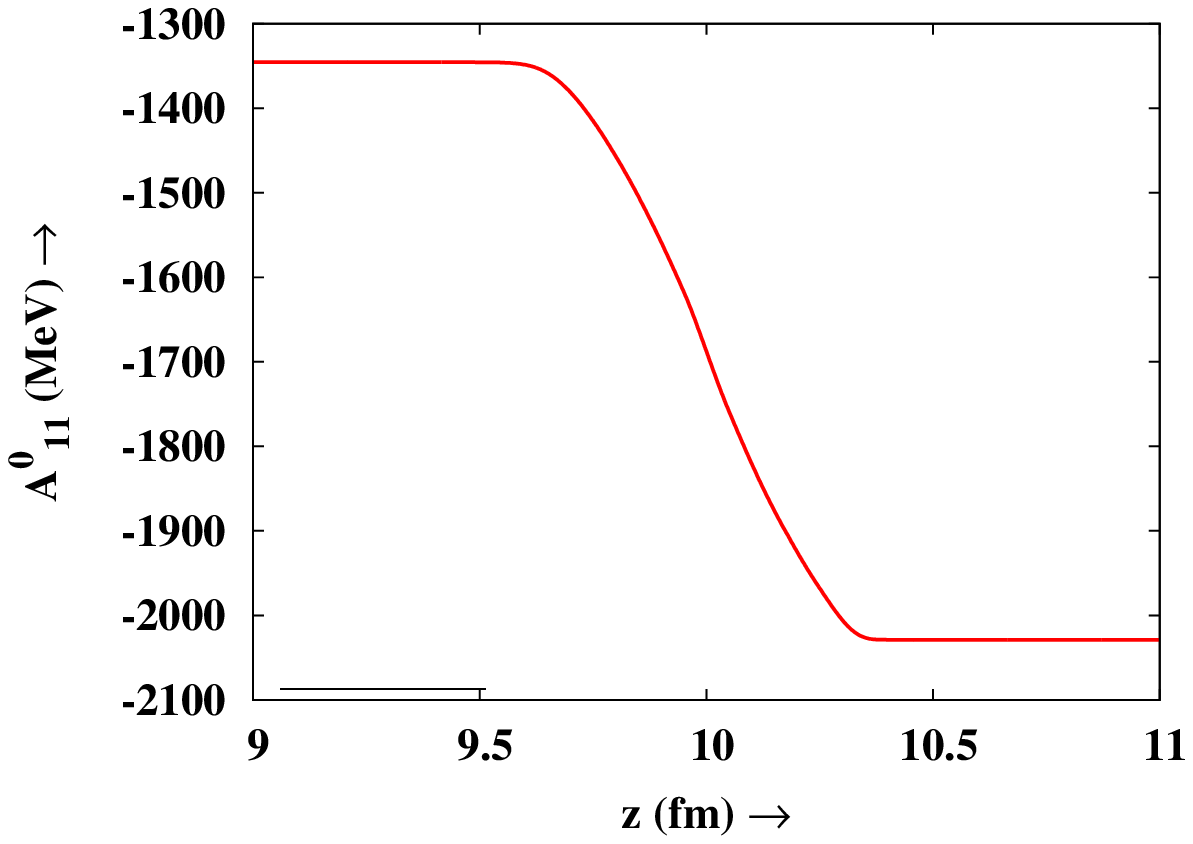} \\
(a) && (b)
\end{tabular} 
\caption{(a) Plot of calculated values of $a$ and $b$ for 
the $|L|$ profile of Fig.4a. (b) corresponding plot of $A_0$.}
\label{fig:a0fkshm}
\end{center}
\end{figure}

  Fig.4a shows the wall profile of $|L|$ for $V[L]$ in Eq.(10) (again,
with normalized $L$). The profile is almost the same as the one
shown in Fig.2a. We mention here that for Fig.4a we have used the
same value of the coefficient of the first $|TrL|^2$ term in Eq.(10)
as with $T_d = 270$ MeV (by suitably changing the values of
string tension etc.). This is so that the shape of the barrier
near the confining vacuum remains unaffected (which determines
the first order nature of the transition). In any case, the overall 
features of the profile of the wall, such as its width and height, 
should depend more on the temperature scale rather than on the 
shape of the barrier for the confining vacuum. To check this, we
also calculate the wall profile of $|L|$ for Eq.(10), but now with 
the value of $T_d = 270$ MeV and $T = 400$ MeV. The comparison of the 
two profiles is shown in Fig.4b. We see that the two profiles are very 
close to each other confirming above arguments.

  We recalculate the plots of $a$ and $b$ for the case with $T = 391$
MeV (with $T_d = 178$ MeV). The resulting plots are shown in Fig.5a
which are seen to be very similar to those on Fig.2b. Finally, the 
profile of $A^0_{11}$ in Fig.5b is also very close to the one in Fig.3. 
Note that though overall all the plots in Figs.(4),(5) are very close
to the corresponding plots in Figs.(2),(3), there is one clear
difference. The profiles in Figs.(4),(5) have somewhat sharper
variations from their asymptotic values compared to the case in
Figs.(2),(3). This originates from the qualitatively different 
shapes of the two potentials in Eq.(5) and Eq.(10) near the region
of Z(3) vacua, and in that sense characterizes the difference in the
two potentials. 

These results are quite remarkable.
Even though the two effective potentials Eq.(5) and Eq.(10) 
(from refs. \cite{Pisarski:2000eq}  and  \cite{Fukushima:2003fw}) are of 
qualitatively different shapes, the resulting wall profile and $A_0$ 
profile are almost the same. As we mentioned above, for small values
of $L$ the two effective potentials will have similar forms, which are
fitted with the Lattice data. Our results thus point out that the profile
of $L$ (and consequently, the profile of $A^0$) are primarily determined
by the small $L$ region of the effective potentials. This is likely
to happen if the variations near the $Z(3)$ vacua are primarily in
the magnitude of $L$ and not in its phase.  The robustness of our results
against different choices of the effective potentials gives us confidence 
in the use of our procedure to calculate the reflection of quark and 
antiquarks from the Z(3) interfaces. Since the $A_0$ profiles of Fig.3 and
Fig.5 are almost the same, the resulting values of reflection coefficients 
for quarks/antiquarks will also be very similar. In the rest of analysis 
in the paper, we will use the effective potential as given in Eq.(5). 

\section{Calculating Reflection and Transmission Coefficients}
\label{sec:ref-trans}

\begin{figure}[!htp]
\begin{center}
\includegraphics[width=0.6\textwidth]{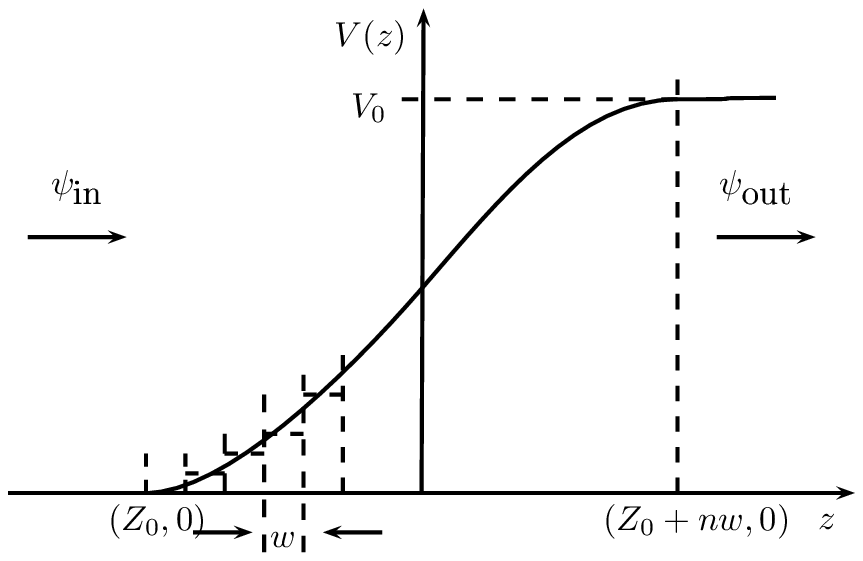}
\caption{Potential ($V(z)$) approximated by a sequence of $n$ step 
potentials, each of width $w$.}
\label{fig:smooth-pot}
\end{center} 
\end{figure}

To calculate the reflection and transmission coefficient, we need the 
solutions of Dirac equation in the Minkowski space. 
We start with the Dirac eqn. in the two dimensional Euclidean space 
\begin{equation}
\label{eq:eucliddiraceq}
\bigl[i\gamma^{0}_{e}\partial_{0}\delta^{jk}  - g\gamma^{0}_{e}A_{0}^{jk}(z) 
+ (i \gamma^{3}_{e}\partial_{3} + m)\delta^{jk}\bigr]\psi_{k} = 0,
\end{equation}
where $\gamma^{0}_{e} \equiv i\gamma^{0}$ and $\gamma^{3}_{e} \equiv 
\gamma^{3}$ are the Euclidean Dirac matrices. $\partial_0$ denotes 
$\partial/\partial_\tau$ with $\tau = it$ being the Euclidean time.
$j,k$ denote color indices.  We now analytically continue
the eqn (\ref{eq:eucliddiraceq}) to the Minkowski space to get
\begin{equation}
\label{eq:minkdiraceq}
\bigl[i\gamma^{0}\partial_{0}\delta^{jk} + g\gamma^{0}A_{0}^{jk}(z) + 
(i\gamma^{3}\partial_{3} + m\bigr)\delta^{jk}]\psi_{k} = 0.
\end{equation}
where now $\partial_0$ denotes $\partial/\partial t$ in the Minkowski
space.
Note that the $A_{0}$ in eqn (\ref{eq:minkdiraceq}), which is in the
Minkowski space, is fundamentally different from the $A_{0}$ in 
eqn (\ref{eq:eucliddiraceq}) which is in the Euclidean space. 
However, it is the \textit{same domain wall profile} (i.e same $A_{0}$ 
dependence on $z$) that appears in both the cases, which is what is 
needed for the calculation of reflection and transmission coefficients. 
For a wave function with time dependence $\psi(x)e^{-iEt}$, the 
eqn (\ref{eq:minkdiraceq}) reduces to
\begin{equation}
\label{eq:tinddiraceq}
\bigl[\gamma^{0}\gamma^{3}\partial_{3}\delta^{jk} + \gamma^{0}m\delta^{jk}\bigr]\psi_{k}(x) = (E - V_{0}(z))\psi_{k}(x).
\end{equation}
where $V(z) = -gA_{0}^{jk}(z)$ is the potential as seen by the incoming 
fermion. We do not have any analytic way to calculate the reflection and 
transmission coefficients for a general smooth potential, so we follow 
a numerical approach. Kalotas and Lee \cite{Kalotas:1991kl} have discussed 
a numerical technique to solve Schr\"{o}dinger equation , approximating
a general smoothly varying (in space) potential in terms of a sequence of 
step functions. We follow their approach and apply their technique for 
solving the Dirac equation (eqn (\ref{eq:tinddiraceq})).
We approximate the actual potential by $n$ step 
potentials in series, each of equal width $w$ as shown in figure 
(\ref{fig:smooth-pot}). 
Let $\psi_{j}$ be the wave-function for the $j^{\textrm{th}}$ bin and the 
height of potential be $V_{j}$. (We consider spin up wave function and
restrict to no-spin-flip situation.) The height of the $j^{\textrm{th}}$ step
potential is taken to be the mean value of $V(L+jw)$ and $V(L+(j+1)w)$, i.e
\begin{equation}
V_{j} = \frac{[V(L+jw) + V(L+(j+1)w)]}{2}
\end{equation}

We now apply boundary conditions at $j^{\textrm{th}}$ step i.e at 
$z = L + jw$. This gives us a set of two equations, which when iteratively 
solved give

\begin{subequations}
\label{eq:ainbin}
\begin{align}
\begin{pmatrix}
 A_{\textrm{in}} \\ B_{\textrm{in}}
\end{pmatrix} &= M^{-1}(L,k_{\textrm{in}})\times M(L,k_{1})\times\dots M^{-1}(L+nw,k_{n})\times M(L+nw,k_{\textrm{out}})\begin{pmatrix}
                                          A_{\textrm{out}} \\ 0
                                         \end{pmatrix} \\
 M(L+jw,k_{q}) &= \begin{pmatrix}
					   e^{ik_{q}(L+jw)} & e^{-ik_{q}(L+jw)} \\
					   \frac{e^{ik_{q}(L+jw)}k_{q}}{E_{q} + m} & -\frac{e^{-ik_{q}(L+jw)}k_{q}}{E_{q} + m}
					  \end{pmatrix}
\end{align}
\end{subequations}
with  $k_{q} = \sqrt{E_{q}^{2} - m^{2}}$, and $E_{q} = E - V_{q}$. 
(Here no left moving wave is allowed in the region far right of the 
interface.) The reflection and transmission coefficients are then given by
\begin{subequations}
\label{eq:reftrans}
\begin{align}
R &\equiv \left\lvert \frac{J_{\textrm{ref}}}{J_{\textrm{in}}}\right\rvert = \left\lvert \frac{B_{\textrm{in}}}{A_{\textrm{in}}}\right\rvert \hspace{4.0cm} \\
T &\equiv \left\lvert \frac{J_{\textrm{trans}}}{J_{\textrm{in}}}\right\rvert =\left\lvert \frac{A_{\textrm{out}}}{A_{\textrm{in}}}\right\rvert\times r \\
\textrm{where} \hspace{0.25cm} r &= \left(\frac{k_{\textrm{out}}}{k_{\textrm{in}}}\right)\left(\frac{E + m}{E - V_{max} + m}\right). 
\end{align}
\end{subequations}

Here, $k_{in} = \sqrt{E^2 - m^2}$ and $k_{out} = \sqrt{(E-V_0)^2 - m^2}$.

\section{Results}
\label{sec:results} 

We first calculated the reflection and transmission coefficients by
assuming the $A_{0}$ profile to be a step function rather than a smooth
one, with the height of the step function being the same as that of the interface in Fig. (\ref{fig:smooth-pot}). In this approximation one can 
calculate the reflection and transmission
coefficients analytically. For anti-quarks the reflection and transmission
coefficients are obtained by changing $g \rightarrow -g$, as anti-quarks are
in $\bar{3}$ representation of $SU\left(3\right)$. We have chosen the 
energies of the particles such that $E > V + m$, so as to avoid the Klein
paradox regime. Note that if $E < V$ (but $V - E < m$ so that one
is away from Klein paradox situation), then the reflection coefficient
for quarks is 1 (repulsive potential) but for antiquarks reflection
coefficient will be very small with $-V$ providing the attractive
potential. This will provide the most dramatic difference between the
reflection of quarks and that of antiquarks from Z(3) walls. However,
for the relevant energies of quarks/antiquarks at RHICE, we discuss
in detail the case with $E > V + m$.

 The results for different quarks and anti-quarks (with $E$ = 3.0 GeV for
each case) are given in table \ref{tab:refquarks}. It is clear that quarks 
have different reflection coefficients than their CP conjugates. Also, the 
effect is significantly higher for the heavier quarks (for example charm quark).
\begin{table}[!htp]
\begin{center}
\begin{tabular}{|c|c|c|c|c|}
\hline
 & $u$ & $d$ & $s$ & $c$ \\
\hline
$E$(GeV) & $3.0$ & $3.0$ & $3.0$ & $3.0$ \\
\hline
$m$(MeV)& $2.5$  & $5.0$ & $100 $ & $1270$ \\
\hline
$R_{q}$ & $1.73 \times 10^{-7}$ & $6.76 \times 10^{-7}$ & $2.8\times 10^{-4}$ & 
$0.14$ \\
\hline 
$R_{\bar{q}}$ & $1.92  \times10^{-8}$ & $7.55 \times 10^{-8}$ & $3.2\times 10^{-5}$ 
& $6.5\times 10^{-3}$\\
\hline
\end{tabular}
\caption{Table for the reflection coefficients for various quarks in
the step function approximation. Reflection is higher for heavier quarks.}
\label{tab:refquarks}
\end{center}
\end{table} 

We now calculate the reflection coefficient for charm quark using the exact 
potential. The product of the matrices in eqn (\ref{eq:ainbin}) were calculated by a FORTRAN code and also by using Mathematica. Eqn 
(\ref{eq:reftrans}) were then used to calculate the reflection coefficient. At 
$E = 3$ GeV, we get $R = 0.0011$ for $c$ quark while for $\bar{c}$ the 
result is $R = 5.24\times 10^{-10}$. As an additional check on the results 
(for the smooth profile), we consider shrinking of the profile of $A_0$
in $z$ direction, and compared the reflection coefficient (for the c quark 
with 3 GeV energy) with the 
step potential result. The results are summarized in Table \ref{tab:result}. 
We see that the numerical results approach the analytical results of
the step function as $A_0$ profile is shrunk along $z$ to better
approximate a step function. This gives us the confidence that our
numerical technique of solving the Dirac equation is reliable.

\begin{table}[!htp]
\begin{center}
\begin{tabular}{|c|c|}
\hline
Shrinking Factor & Reflection Coeff \\
\hline
No shrinking & $0.0011$ \\
\hline
$0.5$ & $0.017$ \\
\hline
$0.05$ & $0.119$ \\
\hline
$0.005$ & $0.123$ \\
\hline
Step Potential & $0.140$ \\
\hline
\end{tabular}
\caption{Table for the reflection coefficients for c quark, with 3 GeV energy,
when the profile is shrunk.  Results approach the step potential as 
the profile gets narrower.}
\label{tab:result}
\end{center}
\end{table} 

 It is clear that if one considers the situation of quarks/antiquarks
coming from right in Fig. (\ref{fig:smooth-pot}) (i.e. approaching the 
domain wall from
the side with $L = Z$) then antiquarks will have larger reflection
coefficients while quarks will have smaller reflection coefficients.
Also we should mention that Eqn (\ref{eq:tinddiraceq}) is solved by using one 
component of $A_{0}$ profile ($A_{0}^{11}$ in this case), which gives us the 
reflection coefficient for one particular color (say red). Reflection
coefficient for other colors will remain the same when SU(3)$_c$ gauge
transformation is applied on the quark as well as on the vector potential. 
However, there is still an ambiguity of starting with different initial 
sets $(a,b)$ (say in the $L = 1$ vacuum). Different sets lead to different 
profiles for $(a,b)$ across the domain wall, thus $A_{0}$ profile depends 
on the initial condition (which, in turn, will lead to different
reflection coefficients for a quark of a given color). 

 As we mentioned earlier, this ambiguity is reasonable in view of the 
fact that we are extracting information about a colored object ($A_0$)
starting from a colorless variable $L(x)$. Thus there is no reason
to expect unique solution for $A_0$ starting from a given $L(x)$ profile,
even in the diagonal gauge where $A_0$ is determined in terms of
real $(a,b)$.  

 For several sets of values of $(a,b)$ we have checked that  different 
choices of $(a,b)$ are related to each other by color transformation. 
We can explain it in the following way: Say we start with $(a_1,b_1)$ for $L$ 
= 1 vacuum and calculate the profile $(a(x),b(x))$ leading to profile 
of $A_0$. Now $A_0^{11}, A_0^{22}, A_0^{33}$ all have different profiles 
and correspond respectively to scattering of red, blue, and green 
quarks respectively, from the given domain wall profile. Now if we 
start with a different set $(a_2,b_2)$ and calculate the profile of $A_0$ 
then we find (for example) that new $A_0^{11}$ is the same as old $A_0^{22}$ 
(where one started with $(a_1,b_1)$) and new $A_0^{22}$ is the same as old 
$A_0^{11}$. This means that $(a_2,b_2)$ set gives same reflection for blue
quark as $(a_1,b_1)$ gives for the red quark. Thus we say
that our different choices of $(a,b)$ amount to considering quarks of
different colors for a given domain wall profile. Or, equivalently,
for the scattering of a fixed color (say red) quark, different
sets $(a,b)$ lead to domain wall profiles carrying different color
information. (We should mention that this holds for many sets
$(a,b)$ we have checked. However, we do not have a general proof
that this should be true for all sets, though it looks very 
likely in view of the above arguments).

\begin{figure}[!htp]
\begin{center}
\begin{tabular}{ccc}
\includegraphics[width=0.45\textwidth]{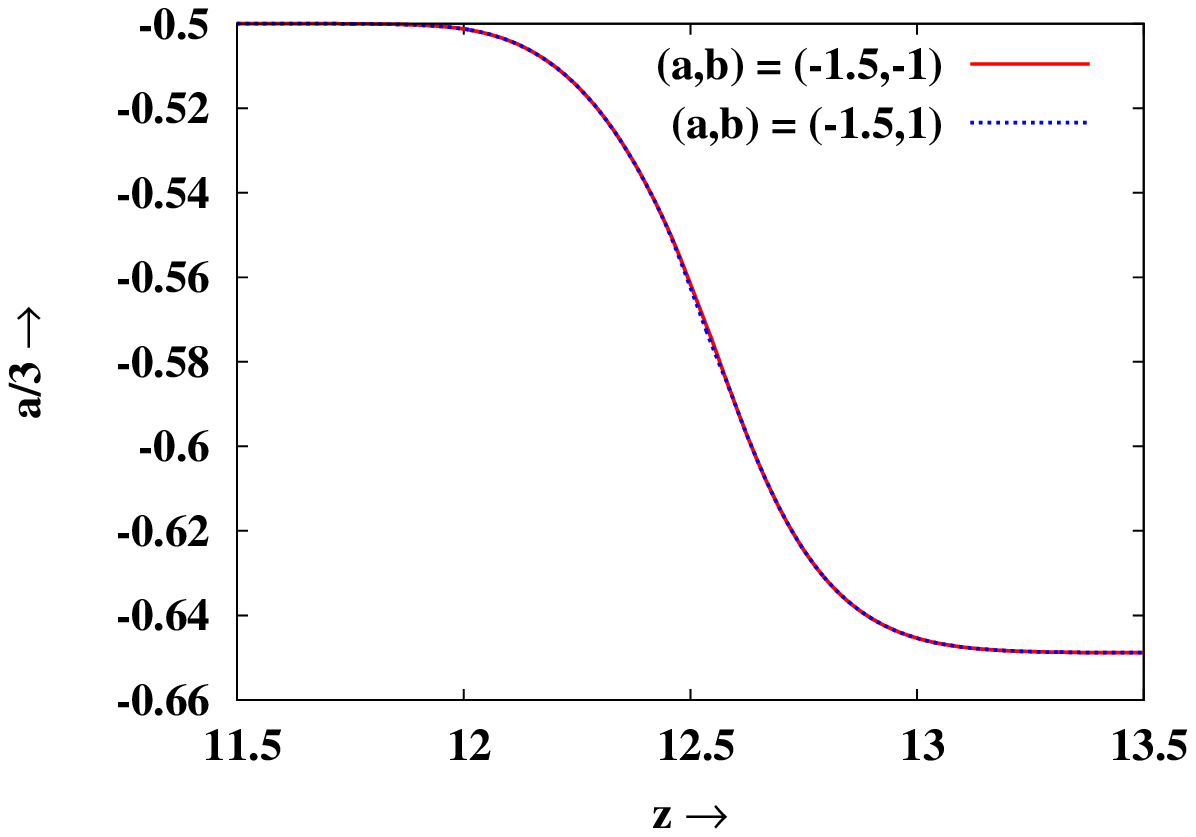}
&\qquad&
\includegraphics[width=0.45\textwidth]{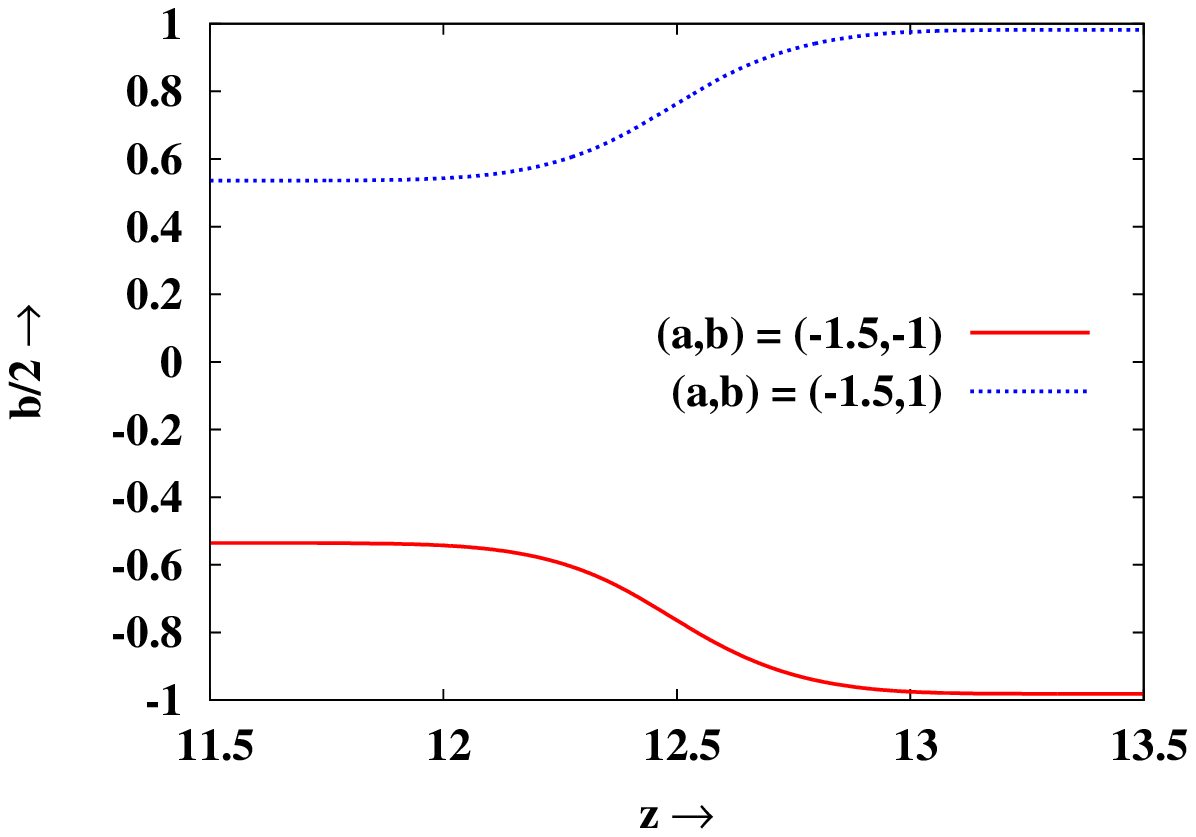} \\
(a) && (b)
\end{tabular} 
\caption{On left: Variation of $a$ for different initial values of  $a,b$. As $a$ is unchanged, it's profile is unaffected. On right: Variation of $b$ for different initial values for $b$. As $b$ changes sign in the initial values, it's profile also changes.}
\label{fig:abvar}
\end{center}
\end{figure}

For example, if we start with $(a^{\prime},b^{\prime}) = (a,-b)$, i.e with 
$(-1.5,1)$, then eqn (\ref{eq:a0diag}) tells us that $A_{0}^{\prime 11} = 
A_{0}^{22}$ and $A_{0}^{\prime 22} = A_{0}^{11}$. See, 
Fig (\ref{fig:abvar}) for 
the corresponding profile of $(a,b)$.
In color space $A_{0}$ is diagonal with elements $(A_{0}^{11}, A_{0}^{22}, 
A_{0}^{33})$, and it acts on the color triplet $(r, b, g)^{T}$. So, 
$A_{0}^{11}$ acting on $(1, 0, 0)^{T}$ is same as $A_{0}^{\prime 22}$ 
acting on $(0, 1, 0)^{T}$ which is same as making different choices 
in color space. 

So, the ambiguity related to various $(a,b)$ profiles or,equivalently, 
corresponding $A_{0}$ profiles, seems to be the artifact of the ambiguity
of making a color choice for the domain wall profile in terms of
$A_0$, starting from the domain wall profile in terms of $L(x)$. 

  This raises an important question whether we should be
dealing with colored domain wall profile (given in terms of
$A_0$ profile) at all, or we should restrict to colorless objects
like $L(x)$ (which is what was done in our earlier works, see ref.4,6). 
After all, the effective potential which we use is given 
in terms of $L(x)$. Here we think that there is no reason to restrict
to colorless objects. We are dealing with the QGP phase and there is
no requirement of physical observables to be color singlets. If we were
dealing with the confining phase then we had obligation of dealing with
colorless objects as physical observables. For QGP phase, it should
make perfect sense to think of domain wall profile having color
properties as it is arising from $A_0$ profile. Of course it is possible
that actual domain wall profile is color insensitive, and quarks of all
colors have same reflection coeff. from a given wall. But it is also
possible that wall is colored and a given wall has different
reflection for quarks of different colors. The only requirement of gauge
invariance is that when color gauge transformations are done on $A_0(x)$
profile as well as on quarks, then numbers should not change, which
is obviously true with the Dirac equation we are using. 

\section{Discussion}
\label{sec:discussion}

This CP violation will have interesting observable consequences for the 
Relativistic Heavy Ion Collision experiments at RHIC and at LHC. If QGP is 
formed in these experiments (and there are strong indications of that), 
then various $Z(3)$ domains will inevitably be formed, leading to the 
formation of Z(3) walls. (We mention that the QGP strings \cite{Layek:2005fn} 
which also necessarily form during transition to QGP phase should also lead to 
spontaneous CP violation. Its effects on quark/antiquarks scattering, or 
possible localization on the QGP strings needs to be explored). 
As these domain walls move/collapse, quarks/anti-quarks will get 
reflected/transmitted differently from these domain walls 
leading to the segregation of quarks and anti-quarks. The concentration 
of quarks (or antiquarks, depending on the collapsing vacuum) will 
grow in different regions of the QGP. As the effects would be stronger 
for heavier quarks (Table \ref{tab:refquarks}), this should lead to 
enhancement of strange and charmed baryons along with the suppression in the
yield of corresponding mesons (such as $J/\psi$). 

Detailed exploration of
the formation and evolution of Z(3) walls and QGP strings in the
context of RHICE has been carried out in ref. \cite{Gupta:2010pp}. 
These simulations show that in the typical region of QGP formed
in RHICE, one expects several Z(3) domain walls to form, their numbers
ranging from 1 to 4,5. The walls may extend throughout the QGP region
with size of order 10 fm. There are  closed domain walls formed
with initial size of about 5-8 fm. The velocities of these walls was
also estimated in ref. \cite{Gupta:2010pp} and were found to range from 
0.5 to 0.8. For detailed discussion of the properties of Z(3) wall and
QGP string networks expected in RHICE, see ref.\cite{Gupta:2010pp}.
These results about the sizes and numbers of Z(3) walls and QGP strings
are very important. This is because one should realize that in a
very large sized QGP region, as in the early Universe, for every domain
wall connecting $\theta=0$ and $\theta=2\pi/3$ vacua, there will be
one connecting $\theta=0$ and $\theta=4\pi/3$ vacua. These walls are
conjugate of each other and the reflection of a quark from the first
wall is identical to the reflection of an antiquark from the second wall.
These two walls are strictly degenerate, even in the presence of 
explicit symmetry breaking effects from dynamical quarks. Thus, on the 
average there will not be any bias for quarks and antiquarks as they
scatter from a network of Z(3) walls.

  This is, however, not true for a small QGP region as produced in
RHICE. As the number of Z(3) walls produced in such a small region
is of order one \cite{Gupta:2010pp}, there may be a net effect 
for the concentration of baryon number, or for anti-baryon, in each 
event. This can be revealed by event-by-event analysis. Even statistically,
for a large number of events, one can calculate the variance of baryon
number density, and spontaneous CP violation from Z(3) walls may be 
detected. For a given event also, segregation of baryons and antibaryons
will occur over large distances of order several fm as indicated by
the typical wall size and separation   \cite{Gupta:2010pp}.

This CP violation can also be very important in the context of early universe 
where it can have interesting implications for generation of baryon
inhomogeneities. As collapsing domain walls preferentially sweep quarks 
(or antiquarks), segregation of quarks and antiquarks will occur.
One can then discuss the formation of baryonic (or antibaryonic) lumps. 
These baryon inhomogeneities can be of large magnitude, with large
separations in the context of certain low energy inflationary models 
\cite{Layek:2005zu}, (but now with CP violation incorporated). We will 
present a detailed study of this in a future work. 

Another important consequence will be on the $P_{t}$ spectra of
hadrons. The quarks/anti-quarks with high momenta will undergo non-trivial
scattering from these Z(3) walls. As $Z(3)$ walls collapse, some get 
transmitted while others are reflected back. For $Z(3)$ walls forming 
closed, collapsing, structures, the quarks
suffer multiple reflections inside the wall, resulting in an increment
in their transverse momenta. This process continues until the walls
either melt away or collapse completely. So the final transverse
momentum of some quarks may be reasonably enhanced before they escape. One
can then use a specific model (such as Recombination/Coalescence model)
to study the $P_{t}$ spectra of final state hadrons, which should show
an increase in the yield of hadrons at high $P_{t}$. 
This has been discussed in ref.\cite{Mishra:2011zz}, however, no account of
CP violation was considered in that work. In the presence of CP violation,
the modified $P_T$ spectra will be different for quarks and for antiquarks. 
We plan to carry out these analyses in a future work.

  The most important limitation of our analysis is the absence of 
quark effects. Dynamical quarks will lead to lifting of degeneracy
between different Z(3) vacua, making $L = 1$ vacuum as the true vacuum
as discussed in refs.
\cite{Pisarski:2000eq,Dumitru:2000in,Dumitru:2002cf,Dumitru:2001bf}.
The one-loop corrections from dynamical quarks have also been discussed in
refs.\cite{Ciminale:2007sr,Fu:2007xc,Costa:2008dp,Marko:2010cd}.  
As we mentioned, recent lattice studies \cite{Deka:2010bc} have provided 
evidence for the existence of such metastable Z(3) vacua. Our analysis
above of calculation of $A_0$ profile and calculation of reflection
coefficients for quarks and antiquarks can be straightforwardly
applied for this non-degenerate case and work is underway on this. 
Apart from affecting the numbers (for reflection coefficients), its
most important effect will be on the evolution of Z(3) wall and QGP
string network, (see ref. \cite{Gupta:2011pp} for a detailed simulation
study of these aspects). However, for the case of RHICE, due to small
length (and time) scales involved, the dynamics of Z(3) walls
is likely to remain dominated by the surface tension effects with
the difference in pressure between different vacua not playing
dominant role for such length scales). Thus the above mentioned features
of effects on hadron spectra due to CP violation may remain qualitatively
true for RHICE.

  However, for the universe the entire issue of formation and evolution
of Z(3) walls crucially depends on the importance of quark effects.
Some discussion of this has been provided in \cite{Layek:2005zu} and we
plan to investigate these issues in future in detail. Most important
issue will be to see whether the spontaneous violation of CP discussed
here can lead to a net separation of baryons and antibaryons in the 
universe which will have observational consequences (e.g. from the
strongly constrained nucleosynthesis, which can be used to constrain
various parameters of the model.)

\section{Acknowledgments}
 We are extremely thankful to Chris Korthals Altes,
A. P. Balachandran, Balram Rai, Pankaj Agrawal, 
Sanatan Digal and Rajarshi Ray for their valuable comments. We would also like 
to thank Uma Shankar Gupta, Ananta P. Mishra, P.S. Saumia, Ranjita Mohapatra, 
Partha Bagchi and Vivek Tiwari for fruitful discussions.


\end{document}